# Gauge invariance of dimension two condensate in Yang-Mills theory.

A.A.Slavnov

Steklov Mathematical Institute,
Gubkina st. 8, GSP-1, 117966 Moscow, Russia, and
Moscow State University

April 23, 2018

**Abstract**

Gauge independence of dimension two vacuum condensate in Yang-Mills theory is demonstrated by using a noncommutative theory technique.

## 1 Introduction

Vacuum condensates are beleived to play an important role in nonperturbative dynamics of gauge fields. The well known example is the dimension four gluon condensate $< \alpha_s \operatorname{tr}(F_{\mu\nu}^2) >$ which proved to be very important in the analysis of the QCD sum rules ([1]). Another possible condensate, having the canonical dimension two, is given by the vacuum expectation value $< \operatorname{tr}(A_\mu^2) >$ Obvious drawback of this condensate is related to the fact that the operator $\operatorname{tr}(A_\mu^2(x))$ is not gauge invariant and it seems that its expectation value, being gauge dependent, cannot have a physical significance. Nevertheless this condensate was used in studies of gauge dependent objects like gluon propagator ([2], [3]). More recently it was argued that nonzero value of $< \operatorname{tr}(A_\mu^2) >$ may be important for the study of the topological structure of vacuum and quark confinement in QCD ([4], [5], [6]). The authors of the papers ([5], [6]) showed that even if $< \operatorname{tr}(A_\mu^2) >$ is not a gauge invariant quantity, its minimal value has a gauge invariant meaning. Further development of these ideas lead to the discovery of BRST-invariant condensate [7])

$$\Omega^{-1} < \int_\Omega d^4x \operatorname{tr}(\frac{1}{2}A_\mu^2(x) - i\alpha c(x)\bar{c}(x)) > \tag{1}$$

where $c(x)$ are Faddeev-Popov ghosts and $\Omega$ is the integration volume. The role of this condensate in the operator product expansion of gauge invariant amplitudes was also discussed.

All these studies strongly suggest that dimension two vacuum condensates are important for the structure of gauge theories vacuum. In this connection a gauge dependence of these condensates seems to be an unfortunate and disturbing phenomenon.

In this paper we show that although the operator $\text{tr}\,(A_\mu^2(x))$ is gauge variant, the expectation value $<\text{tr}\,(A_\mu^2(x))>$ does not depend on a gauge and may have a direct physical meaning. In this proof a technique of noncommutative gauge theories happened to be very useful. It allows to prove gauge independence of dimension two condensate in a simple and elegant way. The explicit construction is presented in the Section 2. This unexpected phenomenon has a natural interpretation if one considers a noncommutative Yang-Mills theory as a matrix model (see [8]). The corresponding model is described in the Section 3. In this section we also mention a possibility of a gauge invariant modification of the Yang-Mills Lagrangian including a "mass" term. It is worth mentioning that using of noncommutative formalism in this paper is mainly a technical tool and our results refer to the usual commutative theory as well.

## 2 Gauge invariant dimension two condensate.

Gauge invariance of the dimension two vacuum condensate may be proven by using a technique of noncommutative gauge theories.

Noncommutative $U(N)$ invariant theory is described by the Yang-Mills type action
$$S = -\frac{1}{8g^2}\int d^4x\,\text{tr}\,(F_{\mu\nu}F_{\mu\nu}) \tag{2}$$
where the curvature tensor is equal to
$$F_{\mu\nu} = \partial_\mu A_\nu - \partial_\nu A_\mu - i[A_\mu * A_\nu - A_\nu * A_\mu] \tag{3}$$
Here $*$ denotes the Moyal product
$$\phi(x) * g(x) = \exp\{\frac{i}{2}\xi\theta^{\mu\nu}\partial_\mu^x\partial_\nu^y\}\phi(x)g(y)|_{y=x} \tag{4}$$
with $\theta^{\mu\nu}$ being a nondegenerate skew-symmetric real matrix, $\theta^2 = -I$, where $I$ is a unit matrix. It is easy to see that in the integals of quadratic forms the Moyal product may be replaced by the usual commutative product. For this reason in the eq.(2) we replaced $F_{\mu\nu} * F_{\mu\nu}$ by $F_{\mu\nu}F_{\mu\nu}$.

The fields $A_\mu$ belong to $U(N)$ algebra: $A_\mu = A_\mu^a t^a$, $a = 0, i$, where $i$ denote $SU(N)$ indices.
$$t^0 = \frac{\sqrt{2}}{\sqrt{N}}I; \quad \text{tr}\,(t^a t^b) = 2\delta^{ab}; \quad [t^a, t^b] = if^{abc}t^c;$$
$$\{t^a, t^b\} = \frac{4}{N}\delta^{ab} + d^{abc}t^c \tag{5}$$

The action (2) is invariant with respect to noncommutative $U(N)$ transformations
$$A_\mu \to \omega * A_\mu * \omega^+ + i\omega * \partial_\mu \omega^+ \tag{6}$$

$$\omega = \exp\{*i\epsilon\} = \sum_n \frac{1}{n!}(i\epsilon)^{*n} \tag{7}$$

It is well known that in the noncommutative case it is impossible to construct a $SU(N)$ invariant action, as $SU(N)$ algebra is not closed under Moyal multiplication. However in the limit $\xi \to 0$ gauge transformations (6) split into the usual $SU(N)$ and $U(1)$ transformations and the Lagrangian becomes the sum of the $SU(N)$ Yang-Mills Lagrangian and the Lagrangian of the free $U(1)$ field.

The limit $\xi \to 0$ in quantum theory is much more tricky. One loop radiative corrections for a finite $\xi$ may be calculated using a standard renormalization technique, however the result is singular at $\xi = 0$. Moreover, the corresponding amplitudes have nonphysical singularities at zero external momenta. The higher order diagrams acquire nonintegrable infrared singularities making the theory inconsistent ([9], [10], [11], [12]).

The situation is quite different if one considers a regularized noncommutative theory. A regularized may be written in the form

$$\Gamma^\Lambda = \sum_i \int f_i^\Lambda(p,k) T_i(p,k) dk_1 \ldots dk_n \tag{8}$$

where $f_i^\Lambda(p,k)$ is a rational function of $p,k$ and $T_i$ is a product of trigonometric functions

$$\sin(\xi p_l^\mu \theta_{\mu\nu} k_m^\nu), \quad \sin(\xi p_l^\mu \theta^{\mu\nu} p_m^\nu), \quad \sin(\xi k_l^\mu \theta^{\mu\nu} k_m^\nu)$$
$$\cos(\xi p_l^\mu \theta^{\mu\nu} p_m^\nu), \quad \cos(\xi p_l^\mu \theta_{\mu\nu} k_m^\nu), \quad \cos(\xi k_l^\mu \theta^{\mu\nu} k_m^\nu)$$

The index $\Lambda$ means that some gauge invariant regularization is introduced, so that

$$\int dk_1 \ldots dk_n |f^\Lambda(p,k)| < \infty \tag{9}$$

Due to the absolute convergence of the integral (8) the function $\Gamma^\Lambda(p)$, defined by the equation (8) has a definite limit at $\xi \to 0$. The limiting theory is the usual commutative $SU(N)$ theory which may be renormalized by adding necessary counterterms. On the contrary, if one firstly takes the limit $\Lambda \to \infty$, renormalization procedure fails and the limit $\xi \to 0$ does not exist. In the following we shall always assume that some gauge invariant regularization is introduced and the limit $\xi \to 0$ is taken in a regularized theory.

We claim that the expectation value

$$< \mathrm{tr}\,(A_\mu(x) A_\mu(x)) > = \Omega^{-1} \int_\Omega < \mathrm{tr}\,(A_\mu(x) A_\mu(x)) > d^4x =$$
$$\Omega^{-1} \int [\exp\{iS_{GF}\} \int_\Omega \mathrm{tr}\,(A_\mu(x) A_\mu(x)) d^4x] dA \tag{10}$$

where $S_{GF}$ denotes a gauge fixed action (2) including necessary Faddeev-Popov ghosts, does not depend on a gauge. In this equation $\Omega$ is a four dimensional integration volume. Due to translational invariance of the action the r.h.s. of eq.(10) does not depend on $\Omega$ and one can take the limit $\Omega \to \infty$.

The crucial observation concerns the existence of the gauge invariant object, including the integral $\int d^4x \, \text{tr}\,(A_\mu(x)A_\mu(x))$. It looks as follows

$$M = \int \text{tr}\,(\frac{1}{2}A_\mu(x)A_\mu(x) - \xi^{-1}A_\mu(x)\theta^{\mu\nu}x_\nu)d^4x \tag{11}$$

Indeed, under the transformation (6) $M$ transforms as follows

$$M \to \int \text{tr}\,(\frac{1}{2}A_\mu * A_\mu + i\omega * A_\mu * \partial_\mu \omega^+ - \frac{1}{2}\omega * \partial_\mu \omega^+ * \omega * \partial_\mu \omega^+$$
$$-\xi^{-1}\omega * A_\mu * \omega^+ * \theta^{\mu\nu}x_\nu - \xi^{-1}i\omega * \partial_\mu \omega^+ * \theta^{\mu\nu}x_\nu) \tag{12}$$

Using the associativity of the Moyal product and unitarity of matrices $\omega$ we have for $\delta M$ the expression

$$\delta M = \int d^4x \, \text{tr}\,(i\omega * A_\mu * \partial_\mu \omega^+ + \frac{1}{2}\partial_\mu \omega * \partial_\mu \omega^+ - i\omega * A_\mu * \partial_\alpha \omega^+ \theta^{\alpha\beta}\theta^{\mu\nu}\delta^{\beta\nu} -$$
$$-\frac{i}{2\xi}\omega * \partial_\mu \omega^+ * \theta^{\mu\nu}\partial_\nu(x^2) = \frac{1}{2\xi^2}\int \text{tr}\,(\omega * x^2 * \omega^+ - x^2)d^4x = 0 \tag{13}$$

Therefore the integral

$$I = \int [\exp\{iS_{GF}\}\Omega^{-1}\int_\Omega d^4x \, \text{tr}\,(\frac{1}{2}A_\mu(x)A_\mu(x) - \xi^{-1}A_\mu(x)\theta^{\mu\nu}x_\nu)]dA_\mu \tag{14}$$

does not depend on a gauge fixing. Indeed, the transition to arbitrary gauge may be done in a standard way by multiplying the integral (14) by "1"

$$1 = \Delta_F(A)\int \delta(F(A^\omega) - B)d\omega \tag{15}$$

changing the variables $A_\mu^\omega \to A_\mu$ and integrating over $B$ with the weight $exp\{i\int \frac{B^2}{2\alpha}d^4x\}$. Due to gauge invariance of the functional $M$ we shall get the representation for the functional $I$ which looks exactly as eq.(14), but with the different gauge fixing terms.

The expectation value $< \text{tr}\,(A_\mu(x)) >$ is zero due to translational invariance of the action and the linear term in eq.(14), singular at $\xi = 0$, vanishes. So we conclude that the value of the dimension two condensate

$$< \text{tr}\,(A_\mu(x)A_\mu(x)) > = \Omega^{-1} < \int_\Omega d^4x \, \text{tr}\,(A_\mu(x)A_\mu(x)) > \tag{16}$$

is gauge independent.

In all these considerations it was assumed that some gauge invariant regularization was introduced. As we discussed above, keeping the regularization fixed one can take a limit $\xi \to 0$. The resulting theory is the standard $SU(N)$ gauge theory plus free $U(1)$ field. It completes the proof of gauge invariance of dimension two vacuum condensate in the commutative $SU(N)$ Yang-Mills theory.

# 3 Yang-Mills theory as a matrix model

The phenomenon described in the previous section has a simple and natural explanation if one uses the connection between noncommutative Yang-Mills theory and the matrix model, described by the action

$$S = -\frac{1}{8g^2} \int \mathrm{tr}\,( A_\mu(x) * A_\nu(x) - A_\nu(x) * A_\mu(x) - \xi^{-2}\theta^{\mu\nu})^2 d^4x \qquad (17)$$

Here $A_\mu$ belong to $U(N)$ algebra, as discussed above. This action is invariant with respect to the transformations

$$A_\mu(x) \to \omega(x) * A_\mu(x) * \omega^+(x) \qquad (18)$$

Being written in terms of the shifted variables

$$A_\mu - \xi^{-1}\theta^{\mu\nu}x^\nu \qquad (19)$$

it describes the $U(N)$ noncommutative Yang-Mills model.

$$S_{sh} = -\frac{1}{8g^2} \int \mathrm{tr}\,( \partial_\mu A_\nu - \partial_\nu A_\mu - i[A_\mu * A_\nu - A_\nu * A_\mu])^2 d^4x \qquad (20)$$

The symmetry transformation (18) in terms of shifted variables looks as follows

$$A_\mu^\omega(x) = \omega(x) * A_\mu(x) * \omega^+(x) + i\omega(x) * \partial_\mu \omega^+(x) \qquad (21)$$

which is nothing but the noncommutative gauge transformation. Note that the shifted action and the gauge transformations (21) are not singular in the limit $\xi \to 0$. In this limit we obtain the usual commutative Yang-Mills theory.

Clearly integrals of the form

$$\int_\Omega \mathrm{tr}\,( A_\mu^2(x))dx; \quad \int_\Omega \mathrm{tr}\,( [A_\mu(x) * A_\mu(x)]^2)dx \qquad (22)$$

are invariant with respect to the transformations (18). Therefore in the shifted theory the integral

$$\frac{1}{2}\int_\Omega \mathrm{tr}\,( A_\mu(x) - \xi^{-1}\theta^{\mu\nu}x_\nu)^2 d^4x$$
$$= \frac{1}{2}\int_\Omega \mathrm{tr}\,( A_\mu^2 - 2\xi^{-1}A_\mu\theta^{\mu\nu}x_\nu + \xi^{-2}x^2)d^4x \qquad (23)$$

is gauge invariant. The last term is a constant which may be omitted without breaking the gauge invariance. Hence the functional

$$M = \int_\Omega \mathrm{tr}\,( \frac{1}{2}A_\mu(x)^2 + \xi^{-1}A_\mu(x)\theta_{\mu\nu}x_\nu)d^4x \qquad (24)$$

is invariant with respect to the gauge transformation (21).

The construction presented above allows also to write a gauge invariant modification of the Yang-Mills Lagrangian, which includes a "mass term" (the term quadratic in $A_\mu$ without derivatives). Indeed the action

$$S = -\frac{1}{8g^2} \int \{ \, \mathrm{tr} \, ( \, \partial_\mu A_\nu - \partial_\nu A_\mu + i[A_\mu * A_\nu - A_\nu * A_\mu])^2 +$$

$$m^2 \, \mathrm{tr} \, ( \, \frac{1}{2} A_\mu^2 - \xi^{-1} A_\mu \theta^{\mu\nu} x^\nu ) \} d^4 x \qquad (25)$$

is obviously gauge invariant. However a naive limit $\xi \to 0$ is singular. Moreover the action contains the linear term, breaking explicitly translational invariance, so it's physical interpretation is not straightforward. These problems require further investigation.

## 4  Discussion

In this paper we proved that the value of the dimension two vacuum condensate $< \mathrm{tr} \, ( \, A_\mu(x) A_\mu(x)) >$ does not depend on a gauge. The gauge invariance of the condensate follows from a hidden symmetry of Yang-Mills theory, which becomes explicit if one considers it as a limit of the noncommutative gauge model. This symmetry has a simple and natural interpretation in terms of a matrix model.

**Acknowledgements.**
This work was supported in part by Russian Basic Research Foundation under grant 02-01-00126, grant for support of leading scientific schools and the RAS program "Theoretical mathematics".